\documentclass[11pt]{article}

\usepackage[utf8]{inputenc}
\usepackage[T1]{fontenc}
\usepackage{lmodern}
\usepackage{amsmath,amssymb}
\usepackage{booktabs}
\usepackage{longtable}
\usepackage{array}
\usepackage{calc}
\usepackage{enumitem}
\usepackage{microtype}
\usepackage{xcolor}
\usepackage{xurl}
\usepackage[hidelinks]{hyperref}
\usepackage[margin=0.9in]{geometry}
\usepackage{setspace}
\usepackage{etoolbox}

\DeclareUnicodeCharacter{00D7}{\ensuremath{\times}}
\DeclareUnicodeCharacter{2265}{\ensuremath{\geq}}
\DeclareUnicodeCharacter{2264}{\ensuremath{\leq}}
\DeclareUnicodeCharacter{2260}{\ensuremath{\neq}}
\DeclareUnicodeCharacter{2248}{\ensuremath{\approx}}
\DeclareUnicodeCharacter{03BA}{\ensuremath{\kappa}}
\DeclareUnicodeCharacter{2013}{--}
\DeclareUnicodeCharacter{2014}{---}
\DeclareUnicodeCharacter{2018}{`}
\DeclareUnicodeCharacter{2019}{'}
\DeclareUnicodeCharacter{201C}{``}
\DeclareUnicodeCharacter{201D}{''}
\DeclareUnicodeCharacter{2212}{-}
\DeclareUnicodeCharacter{00A0}{~}

\hypersetup{colorlinks=true,linkcolor=black,citecolor=black,urlcolor=blue!50!black,pdftitle={The Substitution Escrow Threshold},pdfauthor={Amadeus Brandes}}
\setlist[itemize]{leftmargin=1.25em,itemsep=0.15em,topsep=0.2em}
\setlist[enumerate]{leftmargin=1.5em,itemsep=0.15em,topsep=0.2em}

\title{\textbf{The Substitution Escrow Threshold: When ``Compatible With'' Becomes Safe Enough to Buy}}
\author{Amadeus Brandes\\
Independent Researcher, Germany\\
\texttt{brandesamadeus@gmail.com}}
\date{July 2026}

\begin{document}
\maketitle
\thispagestyle{plain}
\begin{abstract}
Enterprise infrastructure buyers routinely evaluate compatibility claims---``S3-compatible,'' ``PostgreSQL-compatible,'' ``OpenAI compatible''---as proxies for future substitution options. Yet most compatibility claims do not escrow the substitution path they imply. This paper introduces the Substitution Escrow Threshold, a five-condition framework that determines when a compatibility claim genuinely reduces institutional risk versus merely reducing first-integration cost. The five conditions---boundary closure, executable conformance, custody independence, state and operations reversibility, and extension quarantine---are applied to five infrastructure cases (OCI, Kubernetes, OpenTelemetry, S3, PostgreSQL) that populate five distinct outcome cells. The framework produces actionable diagnostics for enterprise architects, platform engineers, procurement teams, and investors evaluating compatibility-dependent infrastructure decisions, and identifies AI infrastructure as the framework\textquotesingle s most urgent next application.
\end{abstract}

OCI images, Kubernetes conformance, OpenTelemetry instrumentation, S3-compatible object storage, and PostgreSQL-compatible distributed databases all appear to answer the same institutional demand: reduce lock-in without forcing the buyer to abandon modern infrastructure.

Yet enterprises do not treat these claims the same way.

OCI compatibility can become a baseline artifact-admission requirement. Kubernetes conformance is necessary but rarely sufficient for managed-service portability. OpenTelemetry creates a narrow but real escape hatch at the instrumentation and telemetry-export layer. S3 compatibility is operationally valuable but usually requires workload profiling. PostgreSQL compatibility accelerates database evaluation but does not, by itself, settle production substitutability.

The apparent pressure is the same: institutions want portability, optionality, and reduced vendor exposure. The outcomes diverge.

The analytical question is therefore not whether a technology is open, standard, API-compatible, vendor-neutral, or widely adopted. The question that matters to a senior platform leader, enterprise architect, security reviewer, procurement team, foundation executive, or infrastructure investor is narrower:

When does compatibility become institutional risk reduction?

The answer is not ``when the API is open.'' It is not ``when the project is under a foundation.'' It is not ``when many vendors support it.'' It is not even ``when migration is technically possible.''

Compatibility becomes institutional risk reduction only when it escrows a future substitution path. A compatibility claim deserves procurement credit only to the extent that the buyer can identify the governed boundary, test the claim independently, rely on custody outside the vendor being constrained, move the relevant state and operating procedures, and quarantine extensions that fall outside the portable core.

That is the \textbf{Substitution Escrow Threshold}.

\section{1. The Observed Puzzle: Same Promise, Different Institutional Weight}\label{the-observed-puzzle-same-promise-different-institutional-weight}

The cases are familiar, but the divergence is not fully explained by the usual language.

OCI is the cleanest case of compatibility becoming procurement-grade. The Open Container Initiative describes itself as an open governance structure for open industry standards around container formats and runtimes, and it contains runtime, image, and distribution specifications. Its certification process requires product testing, publication of test process and results, peer verification, and certification if the product passes. That makes the compatibility claim narrow, governed, and evidence-producing. {[}1{]}{[}2{]}

Kubernetes has a strong conformance program, but the institutional result is different. CNCF says Kubernetes software conformance ensures required API support, enables interoperability between Kubernetes installations, and allows vendors to submit conformance testing results for CNCF review and certification. That supports a real portability claim at the required API core. It does not prove that a production environment can move across managed Kubernetes services without work on identity, storage, networking, ingress, add-ons, observability, upgrade process, and operating model. {[}3{]}

OpenTelemetry sits in another position. It is a vendor-neutral framework for instrumenting, generating, collecting, and exporting telemetry data, and its components include APIs, SDKs, OTLP, semantic conventions, and a Collector that can receive, process, and export telemetry to one or more backends. CNCF lists OpenTelemetry as a CNCF project. That can reduce instrumentation and telemetry-export lock-in. It does not by itself escrow dashboard semantics, query

languages, alert logic, retention policy, incident workflows, or historical-data migration. {[}4{]}{[}5{]}{[}6{]}

S3 compatibility is different again. Amazon publishes the S3 API Reference for actions and data types. {[}7{]} Ceph documents a RESTful API compatible with the basic data-access model of Amazon S3 while also documenting feature support differences, including partial bucket replication and different canned ACL behavior. {[}8{]} MinIO documents the S3 APIs supported by MinIO AIStor and sends readers to Amazon S3 documentation for reference behavior. {[}9{]} This supports a strong practical claim: S3 compatibility can reduce rewrite risk for workloads inside a tested object-operation profile. It does not establish an OCI or Kubernetes-style neutral conformance regime in the cited public record.

PostgreSQL compatibility has still another structure. CockroachDB says it supports the PostgreSQL wire protocol and the majority of PostgreSQL syntax, but also documents unsupported PostgreSQL features and behavioral differences, including cases requiring expression changes. {[}10{]} YugabyteDB says API compatibility is aimed at accelerating developer onboarding and is not aimed at lift-and-shift porting of existing applications. {[}11{]} Its PostgreSQL migration guide frames migration as a transition from monolithic PostgreSQL to YugabyteDB\textquotesingle s distributed architecture requiring planning, schema transformation, data migration, and application optimization. {[}12{]}

The puzzle is not that some claims are true and others false. Most are true at some layer. The puzzle is that the same word - compatibility - carries different institutional meanings depending on where the boundary sits, who governs it, whether it can be tested, what state must move, and how quickly extensions leak outside the portable core.

\section{2. Why the Standard Framings Fail}\label{why-the-standard-framings-fail}

The first failed framing is \textbf{formal standard versus proprietary API}.

It predicts that formal standards should be safe and vendor-originated interfaces should be risky. The cases do not support that cleanly. OCI began from a vendor-shaped container ecosystem but became institutionally stronger because the relevant artifact and runtime boundary moved into open governance and executable certification. S3 compatibility is also vendor-originated and highly useful, but it remains a different kind of claim because third-party implementations document compatibility against Amazon\textquotesingle s reference API rather than against a neutral conformance program comparable to OCI or Kubernetes in the cited public record. {[}1{]}{[}2{]}{[}7{]}{[}8{]}{[}9{]}

The second failed framing is \textbf{foundation governance equals portability}.

Foundation custody matters, but only if the governed boundary is the boundary the buyer depends on. Kubernetes conformance gives real assurance around required APIs. It does not escrow every managed-service dependency that

appears in a production platform environment. OpenTelemetry has a vendor neutral telemetry boundary and is a CNCF project. It does not make every observability backend workflow portable. Governance is necessary for some kinds of compatibility escrow, but it is not sufficient unless the governed surface matches the institutional dependency. {[}3{]}{[}4{]}{[}5{]}{[}6{]}

The third failed framing is \textbf{API compatibility equals substitutability}.

This is the most common procurement error. An API can reduce integration cost without creating an exit right. CockroachDB and YugabyteDB both provide useful PostgreSQL compatibility claims, but their own documentation distinguishes protocol, syntax, drivers, and onboarding from full production migration. A production database workload depends on schema, extensions, query behavior, transaction behavior, distribution, backup and restore, operational runbooks, performance envelope, and failure semantics. The protocol is part of the dependency, not the dependency itself. {[}10{]}{[}11{]}{[}12{]}

The fourth failed framing is \textbf{adoption equals standardization}.

S3 compatibility is widely visible and operationally important. That does not make it the same kind of institutional object as OCI certification or Kubernetes conformance. Widespread implementation can reduce rewrite risk, but procurement-grade portability still requires a known boundary and evidence that the target implementation supports the workload\textquotesingle s actual profile. {[}7{]}{[}8{]}{[}9{]}

The repeated mistake is treating compatibility as a property of software. For institutional buyers, compatibility is a property of a future decision: can the organization later change supplier, runtime, backend, region, deployment model, or operating team without converting the move into a bespoke migration program?

\section{3. Scope}\label{scope}

The Substitution Escrow Threshold applies to infrastructure compatibility claims used to reduce institutional exposure across implementations, vendors, deployment models, or managed services. It covers container artifacts, orchestration APIs, telemetry pipelines, object-storage APIs, database compatibility claims, package infrastructure, AI-serving APIs, developer platforms, and similar infrastructure control points.

It does not apply to ordinary SDK convenience, one-off adapters, UI-level import/export, application-level file exchange, generic benchmark claims, or cases where the buyer is not using compatibility as a risk-reduction argument.

The framework also does not claim that compatibility is the only reason to adopt a system. A buyer may rationally choose a proprietary managed service because it is better operated, cheaper, safer, strategically aligned, or commercially preferable. The framework answers one narrower question:

How much procurement credit should a compatibility claim receive as a future exit right?

\section{4. The Parent Mechanism}\label{the-parent-mechanism}

The parent mechanism is familiar: stable interfaces reduce dependency by

separating one layer from another. Software infrastructure has long relied on

modularity, standards, conformance testing, and interface discipline.

The institutional version is more specific. A technical interface becomes

procurement-relevant when it moves part of the buyer\textquotesingle s future substitution path

outside the vendor\textquotesingle s discretionary control. The buyer no longer has to trust the

vendor\textquotesingle s claim that migration will be possible later. The buyer can point to a

governed boundary, a conformance test, a custody structure, a portable artifact,

a repeatable profile, or a quarantined extension ledger.

Compatibility, in that condition, functions like escrow. The buyer has not

exercised the exit option today. But enough of the route is held outside the

vendor that architecture, security, procurement, and risk teams can assign it

value.

\section{5. Related Work}\label{related-work}

The Substitution Escrow Threshold draws on several literatures that have largely run on separate tracks.

The first is the economics of compatibility and standards. Network-effects theory established why compatibility carries economic weight: the value of committing to an interface rises with the installed base behind it, and a market can lock onto an inferior standard through ``excess inertia'' once expectations converge. {[}25{]}{[}26{]} The path-dependence tradition sharpened this into the claim that small historical events can lock a market onto a standard that endures without being efficient. {[}27{]}{[}28{]}{[}29{]} That tradition also drew the distinction this paper leans on most heavily --- between de jure standards governed through an independent process and de facto standards that emerge from a dominant implementation. {[}30{]}{[}31{]} The framework's ``escrowed'' versus ``borrowed'' compatibility is that distinction restated in procurement terms: OCI as governed conformance, S3 as dominant-implementation compatibility. The vocabulary of lock-in and switching cost that runs through the five conditions is likewise inherited from the information-economics branch of this literature. {[}32{]} The unifying treatment of switching costs and network effects as the joint source of lock-in is the reference point here. {[}33{]}

What that literature does not supply is a buyer-side decision rule. It explains why standards form and how lock-in operates at the level of markets; it does not tell a procurement team when a particular vendor's ``compatible-with'' claim should be credited as a governed exit option rather than a convenience. The five conditions operationalize exactly that gap.

The second literature is the empirical software-engineering work on interface stability. Large-scale studies of semantic versioning find that a substantial share of releases labeled compatible nonetheless introduce breaking changes, and that adherence to versioning conventions varies across package ecosystems, so declared version semantics are an unreliable proxy for what actually moves without rework. {[}34{]}{[}35{]} That is the intra-ecosystem form of Condition 1: a compatibility label reduces risk only to the degree that the production behaviors a workload depends on sit inside the tested boundary. The Substitution Escrow Threshold generalizes the observation from library-and-client co-evolution within a single ecosystem to substitution across vendors, deployment models, and managed services, where the boundary is institutional as well as technical.

The framework also speaks to the information-systems literature on IT sourcing and vendor lock-in, where its intended audience sits. Reviews spanning two decades of IT-outsourcing research treat vendor dependence and switching cost as central determinants of sourcing outcomes, {[}36{]} and survey work on cloud migration finds that lock-in intensifies as workloads move from on-premise to managed services, driven precisely by the absence of enforceable standardization. {[}37{]} That literature establishes vendor lock-in as a first-order procurement risk but largely stops at identifying it; it does not give a buyer a test for when a specific compatibility claim converts that risk into a credible exit option. The five conditions are intended as that test. Buyer-side frameworks for assessing lock-in and selecting portable services do exist, but they operate at the granularity of a service or a migration program; the object this paper scores is narrower --- an individual ``compatible-with'' claim, and how much exit credit it earns. The mechanism the framework prices is an option, not a discount: an escrowed compatibility claim carries the value of an unexercised right to switch under uncertainty, the logic the information-systems literature on real options developed for IT investment and platform-adoption decisions. {[}38{]}{[}39{]} The five conditions specify when that option is actually held rather than merely asserted.

The paper's organizing metaphor should also be distinguished from literal software escrow. Source-code escrow is an established contractual practice --- set against the statutory backdrop of the Intellectual Property Bankruptcy Protection Act {[}40{]} --- in which a vendor's source code is released to the licensee on a defined trigger such as bankruptcy or abandonment, protecting continuity when a vendor fails. Compatibility escrow in the sense used here is different: a standing substitution path the buyer can exercise at will, including when the vendor is healthy but the relationship has simply become disadvantageous. The two are complementary hedges against the same exposure --- dependence on a single supplier --- but they are triggered by opposite conditions: vendor death versus buyer exit.

These questions are no longer only commercial; a regulatory and standards apparatus is now shaping the same procurement decisions. ISO/IEC 19941 supplies shared terminology for cloud interoperability and portability. {[}41{]} A consumer-facing analogue predates both: the GDPR\textquotesingle s Article~20 established a right to data portability for personal data, though it binds data controllers toward data subjects rather than infrastructure buyers. {[}42{]} The EU Data Act goes considerably further, mandating by regulation much of the exit right this paper derives by procurement discipline: its Chapter VI obliges providers of data processing services to remove the obstacles that inhibit switching and, for infrastructure services, to enable ``functional equivalence'' in the destination environment. {[}43{]} This is the closest legal analogue to the Substitution Escrow Threshold, and the contrast is instructive. A mandate that a provider enable switching is not an adjudication of whether any particular ``compatible-with'' claim actually delivers it for a given workload; the regulation creates the right but leaves the buyer to determine whether a specific service clears the bar. The five conditions are that determination --- the buyer-side test a portability mandate presupposes but does not itself supply.

\section{6. The Framework: Five Conditions}\label{the-framework-five-conditions}

A compatibility claim crosses the Substitution Escrow Threshold when it

satisfies five conditions strongly enough that an institution can treat it as a

governed exit option rather than a vendor assertion.

\begingroup\small
\begin{longtable}[]{@{}
  >{\raggedright\arraybackslash}p{(\columnwidth - 4\tabcolsep) * \real{0.30}}
  >{\raggedright\arraybackslash}p{(\columnwidth - 4\tabcolsep) * \real{0.70}}@{}}
\toprule\noalign{}
\begin{minipage}[b]{\linewidth}\raggedright
Condition
\end{minipage} & \begin{minipage}[b]{\linewidth}\raggedright
Operational definition
\end{minipage} \\
\midrule\noalign{}
\endhead
\bottomrule\noalign{}
\endlastfoot
Boundary closure & The interface specification covers the production behaviors the workload relies on. \\
Executable conformance & The claim can be tested or certified by a party other than the vendor. \\
Custody independence & The contract is governed outside the unilateral control of the vendor whose power it is meant to constrain. \\
State and operations reversibility & The buyer can move the relevant state, configuration, policy, and operating procedures within a bounded program. \\
Extension quarantine & Proprietary or implementation-specific features can be identified, isolated, and labeled as departures from the portable contract. \\
\end{longtable}
\endgroup

The first two conditions are gates. Without boundary closure and executable

conformance, compatibility remains mostly assertion. Custody independence

determines whether the claim can survive vendor conflict. State and operations reversibility determines whether the claim reduces actual migration cost.

Extension quarantine determines whether the claim remains durable after real

adoption.

\subsection{6.1 Condition 1: Boundary Closure}\label{condition-1-boundary-closure}

Boundary closure asks whether the compatibility claim covers the behavior that

matters.

A boundary can be narrow and still valuable. OCI does not promise full

application-platform portability. It specifies container image, runtime, and distribution boundaries. {[}1{]} OpenTelemetry does not promise observability-backend

interchangeability. It specifies instrumentation, telemetry APIs, SDKs, OTLP, semantic conventions, and collector behavior. {[}4{]}{[}5{]} Kubernetes conformance does not promise cloud-provider neutrality across every managed-service dependency. It covers required Kubernetes APIs. {[}3{]}

Boundary closure fails when the declared interface covers only a small part of the production dependency. PostgreSQL wire-protocol compatibility is valuable, but a production PostgreSQL environment may depend on extensions, advisory locks, stored procedures, query-planner behavior, replication assumptions, backup tools, migration tooling, operational runbooks, and performance envelopes. CockroachDB\textquotesingle s compatibility documentation is useful precisely because it distinguishes supported PostgreSQL protocol and syntax from unsupported features and behavioral differences. {[}10{]}

The operational test is simple: list the top twenty behaviors the workload relies on. If most sit inside the compatibility contract, boundary closure is high. If most sit in implementation details, managed-service integrations, extensions, undocumented semantics, or operational tooling, boundary closure is low.

\subsection{6.2 Condition 2: Executable Conformance}\label{condition-2-executable-conformance}

Executable conformance asks whether compatibility can be tested by someone other than the vendor.

Documentation is not enough. A compatibility matrix is not enough. A sales claim is not enough. A compatibility claim becomes institutionally useful when there is a versioned test suite, certification process, or reproducible validation procedure that a buyer, auditor, foundation, competing vendor, or internal platform team can run.

OCI and Kubernetes are strong here. OCI certification requires testing, publication of test process and results, peer verification, and certification if the product passes. CNCF\textquotesingle s Kubernetes conformance program requires vendors to submit conformance testing results for review and certification, and CNCF describes the program as ensuring required API support. {[}2{]}{[}3{]}

Executable conformance does not prove that a product is operationally good. It proves that the compatibility claim has an inspectable evidence layer. That is enough to change procurement behavior. A platform team can attach conformance evidence to an architecture-review packet. A security team can require it as an admission control. A procurement team can make it a threshold requirement.

The operational test: can the buyer attach a test result, certification record, or repeatable conformance run to the decision file? If not, the claim should not receive full procurement-grade exit credit.

\subsection{6.3 Condition 3: Custody Independence}\label{condition-3-custody-independence}

Custody independence asks who controls the compatibility contract.

A compatibility claim has weak institutional value if the same vendor whose power it is meant to constrain controls the definition, tests, trademark, interpretation, and change process. Independent custody does not require perfect neutrality. It requires that the contract not be reducible to one provider\textquotesingle s roadmap.

OCI describes an open governance structure for standards around container formats and runtimes. {[}1{]} Kubernetes conformance is run through CNCF. {[}3{]} OpenTelemetry defines vendor-neutral APIs, SDKs, OTLP, semantic conventions, and collector behavior, and CNCF lists OpenTelemetry as a CNCF project. {[}4{]}{[}5{]}{[}6{]} Those structures do not eliminate all risk, but they move the relevant compatibility contract outside unilateral vendor control.

S3 compatibility occupies a different position. The reference API is Amazon\textquotesingle s. Third-party systems such as Ceph and MinIO document compatibility against Amazon S3 behavior or supported S3 APIs. This can still be materially useful, but the cited public record does not establish an OCI- or Kubernetes-style neutral conformance process for S3-compatible implementations. {[}7{]}{[}8{]}{[}9{]}

The operational test: if the dominant vendor changes direction, can the compatibility contract continue to evolve through a credible process controlled by multiple interested parties? If not, compatibility may still reduce integration cost, but it should receive less credit as vendor-exposure reduction.

\subsection{6.4 Condition 4: State and Operations Reversibility}\label{condition-4-state-and-operations-reversibility} State and operations reversibility asks whether real production assets can move.

Call-level compatibility is not enough if accumulated state, configuration, identity policy, incident procedures, monitoring, deployment topology, cost assumptions, and runbooks cannot move. This is why Kubernetes conformance and Kubernetes portability are different claims. A cluster may be conformant at the API core while still depending heavily on a cloud provider\textquotesingle s IAM model, load balancers, storage classes, managed upgrades, node lifecycle, ingress behavior, observability integrations, and policy controls. CNCF\textquotesingle s conformance claim supports required API interoperability; the broader managed-service perimeter is an institutional inference outside that conformance boundary. {[}3{]}

OCI scores strongly inside its boundary because the portable object is the artifact itself: the OCI site describes runtime, image, and distribution specifications, and OCI certification includes runtime and image certification programs. That supports artifact/runtime portability, not full platform portability. {[}1{]}{[}2{]}

OpenTelemetry scores strongly for telemetry emission and instrumentation decoupling because its documentation covers APIs, SDKs, OTLP, semantic conventions, and a vendor-agnostic Collector. It scores less strongly for backend workflow because the OpenTelemetry boundary does not itself standardize dashboards,

alerts, query languages, retention policy, incident process, or historical-data migration. {[}4{]}{[}5{]}

S3 compatibility may move object data more easily than database state, but production object-storage dependency frequently includes lifecycle rules, object lock, event notifications, encryption modes, replication behavior, access policy, and cost behavior. Ceph\textquotesingle s own S3 documentation shows why this must be profile-specific: support is not merely a binary label, and some features are partial or behaviorally different. {[}8{]}

PostgreSQL-compatible distributed databases face the hardest version of this problem because schema, queries, extensions, transaction behavior, distribution, and operational practices are all part of the production state. YugabyteDB\textquotesingle s migration guide frames migration as planning, schema transformation, data migration, and application optimization for a distributed environment, which supports treating substitutability as workload-specific rather than automatic. {[}12{]}

The operational test: can the buyer move not only code, but the relevant state and operating procedures, within a bounded program? If the answer requires workload-by-workload archaeology, reversibility is low.

\subsection{6.5 Condition 5: Extension Quarantine}\label{condition-5-extension-quarantine}

Extension quarantine asks whether non-portable features can be kept visible.

Most lock-in does not happen at adoption. It accumulates through useful extensions. Teams begin inside the portable subset, then adopt managed identity integration, accelerated storage, proprietary policy controls, backend-specific dashboards, custom operators, lifecycle rules, advanced indexing, eventing integrations, provider-specific metrics, or specialized billing constructs.

Each decision is locally rational. The aggregate effect is to move the deployment outside the escrowed boundary.

A compatibility framework that cannot quarantine extensions decays over time. The question is not whether extensions exist. They always will. The question is whether the institution can label them as departures from the portable contract.

The operational test: can the buyer produce an inventory that separates ``inside compatibility escrow'' from ``provider extension''? If not, the organization may be consuming portability rhetoric while building a non-portable system.

\section{7. The Decision Structure}\label{the-decision-structure}

The framework produces five institutional outcomes.

\begingroup\small
\begin{longtable}[]{@{}
  >{\raggedright\arraybackslash}p{(\columnwidth - 4\tabcolsep) * \real{0.30}}
  >{\raggedright\arraybackslash}p{(\columnwidth - 4\tabcolsep) * \real{0.70}}@{}}
\toprule\noalign{}
\begin{minipage}[b]{\linewidth}\raggedright
Outcome
\end{minipage} & \begin{minipage}[b]{\linewidth}\raggedright
Threshold state
\end{minipage} \\
\midrule\noalign{}
\endhead
\bottomrule\noalign{}
\endlastfoot
Escrowed compatibility & Boundary is closed, conformance is executable, custody is independent, state and operations are reversible, and extensions are quarantined. \\
Governed core, exposed perimeter & Core API is governed and testable, but production perimeter leaks through managed-service and organization-specific dependencies unless separately controlled. \\
Narrow escape hatch & A deliberately narrow layer is standardized and vendor-neutral. \\
Borrowed compatibility & A dominant vendor-originated API is widely implemented but not independently governed or supported by a neutral conformance program. \\
Onboarding compatibility & Protocol or syntax compatibility exists, but semantic state and operations remain workload-specific. \\
\end{longtable}
\endgroup

This is the central claim: \textbf{compatibility changes institutional risk only}

\textbf{when it escrows the buyer\textquotesingle s future substitution path.}

A claim that cannot be bounded, tested, governed, operationally reversed, and

protected from extension drift may still be useful. It may reduce integration

cost. It may accelerate evaluation. It may improve bargaining position. But it

should not materially reduce assessed vendor exposure.

\section{8. Case Matrix}\label{case-matrix}

\begingroup\footnotesize
\begin{longtable}[]{@{}
  >{\raggedright\arraybackslash}p{(\columnwidth - 14\tabcolsep) * \real{0.11}}
  >{\raggedright\arraybackslash}p{(\columnwidth - 14\tabcolsep) * \real{0.15}}
  >{\raggedright\arraybackslash}p{(\columnwidth - 14\tabcolsep) * \real{0.15}}
  >{\raggedright\arraybackslash}p{(\columnwidth - 14\tabcolsep) * \real{0.15}}
  >{\raggedright\arraybackslash}p{(\columnwidth - 14\tabcolsep) * \real{0.15}}
  >{\raggedright\arraybackslash}p{(\columnwidth - 14\tabcolsep) * \real{0.15}}
  >{\raggedright\arraybackslash}p{(\columnwidth - 14\tabcolsep) * \real{0.09}}@{}}
\toprule\noalign{}
\begin{minipage}[b]{\linewidth}\raggedright Case\end{minipage}
 & \begin{minipage}[b]{\linewidth}\raggedright Boundary closure\end{minipage}
 & \begin{minipage}[b]{\linewidth}\raggedright Executable conformance\end{minipage}
 & \begin{minipage}[b]{\linewidth}\raggedright Custody independence\end{minipage}
 & \begin{minipage}[b]{\linewidth}\raggedright State \& operations reversibility\end{minipage}
 & \begin{minipage}[b]{\linewidth}\raggedright Extension quarantine\end{minipage}
 & \begin{minipage}[b]{\linewidth}\raggedright Outcome\end{minipage} \\
\midrule\noalign{}
\endhead
\bottomrule\noalign{}
\endlastfoot
OCI & High: image, runtime, and distribution boundary is narrow and explicit. & High: certification supplies test process, published results, peer verification, and certification if the product passes. & High: the compatibility boundary is independently governed through open industry standards rather than one vendor's implementation. & High within the artifact/\allowbreak runtime/\allowbreak distribution boundary; not a claim of full platform portability. & High: surrounding platform, IAM, storage, service-mesh, secrets, observability, and incident-response dependencies remain outside the escrowed boundary. & Escrowed compatibility \\
Kuber\-netes & Medium-high at required API core; lower at managed-service perimeter. & High at the certified API core: vendors submit conformance testing results for review and certification. & High at the core API level through CNCF-governed conformance. & Medium: core API interoperability is supported, but real deployments accumulate cloud IAM, load balancer, CSI, ingress, secret-store, service-mesh, logging, backup, upgrade, and runbook dependencies. & Medium: extensions can be quarantined only if the buyer maintains a portability ledger separating conformant core, standardized ecosystem dependency, and provider-specific extension. & Governed core, exposed perimeter \\
Open\-Tele\-metry & High at instrumentation and telemetry-transport boundary. & Medium: the layer is specified through APIs, SDKs, OTLP, semantic conventions, and Collector behavior, but the cited prose does not establish an OCI/Kubernetes-style certification program. & High: the relevant layer is vendor-neutral and under CNCF project custody. & Medium: telemetry emission and instrumentation are decoupled, but dashboards, queries, alerts, retention, incident workflow, RBAC, billing behavior, and historical data remain outside the standard boundary. & Medium-high: the narrow boundary makes backend-specific workflow visible as outside escrow, but it does not itself make observability backends substitutable. & Narrow escape hatch \\
S3 compatibility & Medium: strong object API center; broader surrounding behavior varies. & Medium-low: the workload profile can be tested, but the cited record does not establish neutral conformance comparable to OCI or Kubernetes. & Low: the reference interface is AWS-controlled, with third parties documenting support against that interface. & Medium: object data is often more movable than database state, but lifecycle, versioning, object lock, IAM/policy, encryption, replication, eventing, consistency, storage-class, residency, and cost behavior may remain workload-specific. & Medium: required operations, headers, auth flows, lifecycle rules, event integrations, encryption, object-lock semantics, replication, failure behavior, performance, and retention controls can be profiled and tested, but not assumed portable from the label alone. & Borrowed compatibility \\
Post\-greSQL compatibility & Medium-low for production substitution: protocol and syntax are not workload behavior. & Low-medium: compatibility is mainly vendor documentation plus workload assessment, not a neutral executable conformance boundary for production substitution. & Medium-low: PostgreSQL is a broad ecosystem surface, but the relevant distributed-database compatibility claim is still largely vendor-defined and workload-specific. & Low: schema, extensions, planner assumptions, transaction behavior, operational tooling, backup and restore practices, performance envelope, and failure semantics must be validated for the actual workload. & Medium-low: unsupported and partially supported feature lists help expose departures, but production substitution still requires workload-level validation. & Onboarding compatibility \\
\end{longtable}
\endgroup

The matrix is not a ranking of technologies. It is a method for assigning

procurement credit to specific compatibility claims.

The five cases function as an analytic typology, not a validation corpus. They were selected because they make visible five distinct ways a compatibility claim can acquire, or fail to acquire, procurement-grade substitution value. A validation study would require independently coded procurement artifacts, migration records, conformance evidence, RFP language, and workload-profile tests across multiple institutions; Section 11 states what such evidence would need to show to weaken each assignment.

\subsection{8.1 Case 1: OCI - Escrowed Compatibility}\label{case-1-oci-escrowed-compatibility}

OCI is the cleanest example of compatibility becoming procurement-grade

because the boundary is narrow, governed, and testable.

The institutional problem in the container ecosystem was not simply that containers needed a format. It was that enterprises needed the container artifact to

stop being a bet on one vendor\textquotesingle s implementation. OCI converted a key vendor

shaped interface into an independently governed compatibility boundary. The

initiative describes itself as creating open industry standards around container

formats and runtimes and lists runtime, image, and distribution specifications.

Its certification process adds an evidence layer: test the product, publish test

process and results, allow peer verification, and receive certification if the product

passes. {[}1{]}{[}2{]}

This is why OCI compatibility can appear in platform standards, artifact

admission rules, registry requirements, and supply-chain controls without being

empty. A security team can require OCI images. A platform team can require OCI-compatible registries or runtimes. A supply-chain team can build signing, scanning, and provenance processes around OCI artifacts without assuming that one vendor\textquotesingle s registry, image builder, or runtime remains permanent.

The framework also prevents overclaiming. OCI does not make a complete production environment portable. It does not standardize cloud IAM, persistent volumes, service meshes, deployment workflows, runtime policy, secrets management, observability, or incident response. It escrows the artifact/runtime/distribution boundary.

That boundary discipline is the reason the claim is strong. OCI compatibility matters because it does not pretend to be more than it is.

\subsection{8.2 Case 2: Kubernetes - Governed Core, Exposed Perimeter}\label{case-2-kubernetes-governed-core-exposed-perimeter}

Kubernetes crosses the threshold at the core API level, but not at the whole operating-model level.

CNCF states that software conformance ensures that vendors\textquotesingle{} Kubernetes versions support required APIs and that conformance enables interoperability from one Kubernetes installation to another. Vendors submit conformance testing results for review and certification. That is real institutional evidence. It explains why enterprise buyers can standardize around Kubernetes while preserving some choice among distributions, managed services, and platform vendors. {[}3{]}

But certified Kubernetes is not the same as portable infrastructure.

A real Kubernetes production deployment accumulates provider-specific and organization-specific dependencies: cloud IAM, load balancers, CSI drivers, node images, autoscaling behavior, managed control planes, ingress controllers, secret stores, service meshes, policy engines, logging integrations, backup tools, upgrade practices, and incident runbooks. Some of these are standardized. Some are portable with work. Some are tightly bound to a managed provider.

The correct procurement statement is not ``Kubernetes makes us cloud-portable.'' It is:

Kubernetes conformance escrows part of the core control-plane API. The managed-service perimeter remains exposed unless separately controlled.

This has direct consequences for sourcing. A Kubernetes RFP or architecture review should not stop at certified conformance. It should require a portability ledger with three columns: conformant core, standardized ecosystem dependency, and provider-specific extension. The first column can receive exit credit. The second requires validation. The third is vendor exposure and should be priced as such.

Kubernetes did not fail because portability is incomplete. It succeeded because the escrowed boundary is institutionally meaningful. The error is treating the

boundary as larger than it is.

\subsection{8.3 Case 3: OpenTelemetry - Narrow Escape Hatch}\label{case-3-opentelemetry-narrow-escape-hatch}

OpenTelemetry\textquotesingle s strength is not that it makes observability vendors interchange able. Its strength is that it moves instrumentation and telemetry export away from proprietary capture.

The project\textquotesingle s public claim is layer-specific: a vendor-neutral observability framework for instrumenting, generating, collecting, and exporting telemetry data, with components that include APIs, SDKs, OTLP, semantic conventions, and a Collector that can receive, process, and export telemetry data to one or more backends. CNCF lists OpenTelemetry as a CNCF project. {[}4{]}{[}5{]}{[}6{]}

That is materially valuable. A platform team can standardize service instrumentation across languages. It can route telemetry through a collector instead of embedding every service directly into one vendor\textquotesingle s agent path. It can preserve more optionality when changing telemetry destinations.

But the standard does not by itself escrow the full observability workflow. Dashboards, query languages, alert definitions, SLO constructs, retention policies, cardinality controls, incident workflows, RBAC models, billing behavior, and historical data all sit outside the basic instrumentation/export boundary. A company that emits OpenTelemetry data may still face a serious migration program when changing observability backends.

This is not a criticism of OpenTelemetry. It is why the project works. The boundary is narrow enough to be real.

The buyer\textquotesingle s rule should be precise: require OpenTelemetry for new instrumentation, but do not treat OpenTelemetry support as backend substitutability. It reduces lock-in at the telemetry-emission layer. It does not eliminate lock-in at the operational-analysis layer.

OpenTelemetry is the narrow escape hatch: highly useful because it is deliberately bounded.

\subsection{8.4 Case 4: S3 Compatibility - Borrowed Compatibility}\label{case-4-s3-compatibility-borrowed-compatibility}

S3 compatibility is one of the most visible and operationally useful de facto compatibility claims in object-storage procurement. It is also one of the easiest to over-credit.

Amazon publishes the S3 API Reference for actions and data types. {[}7{]} Ceph documents support for a RESTful API compatible with the basic data-access model of Amazon S3 and then publishes feature-support detail, including partial bucket replication and different canned ACL behavior. {[}8{]} MinIO documents the S3 APIs supported by MinIO AIStor and directs readers to Amazon S3 documentation for reference behavior. {[}9{]}

That evidence supports a strong but bounded claim: S3 compatibility reduces integration and rewrite risk for workloads that stay inside the supported object operation profile. It lets applications, backup tools, data platforms, and AI pipelines target a familiar object-storage interaction model. It gives buyers leverage because a large body of software already speaks S3-shaped APIs.

But under the Substitution Escrow Threshold, S3 compatibility is borrowed compatibility rather than escrowed compatibility. The reference interface is controlled by AWS. Third-party implementations document support against that interface. The cited public record does not establish a neutral conformance program comparable to OCI certification or Kubernetes conformance.

The state-reversibility picture is mixed. Object data is often more movable than database state. But production object-storage dependency can include lifecycle policies, versioning, object lock, bucket policy, IAM integration, encryption modes, replication behavior, event notifications, multipart edge cases, consistency expectations, storage classes, data-residency rules, and cost behavior.

The procurement action is not ``accept S3-compatible.'' It is: Define the workload\textquotesingle s S3 profile.

The buyer should list required operations, headers, auth flows, bucket policies, lifecycle rules, event integrations, encryption requirements, object-lock semantics, replication expectations, failure behavior, performance assumptions, and retention controls. Then the buyer should test the target implementation against that profile.

S3 compatibility remains important. It just should not receive general exit-right credit until the relevant workload profile has been tested.

\subsection{8.5 Case 5: PostgreSQL Compatibility - Onboarding Compatibility}\label{case-5-postgresql-compatibility-onboarding-compatibility}

PostgreSQL compatibility is valuable and institutionally limited for the same reason: the surface is deep.

CockroachDB says it supports the PostgreSQL wire protocol and most PostgreSQL syntax, and that existing PostgreSQL applications can often migrate without changing application code. The same documentation says CockroachDB does not support some PostgreSQL features or behaves differently because not all features are easy to implement in a distributed system; it also documents unsupported and partially supported features. {[}10{]}

YugabyteDB\textquotesingle s documentation is even more direct about the institutional boundary. Its compatibility FAQ says API compatibility is aimed at accelerating developer onboarding and is not aimed at lift-and-shift porting of existing applications. Its PostgreSQL migration guide frames migration as a transition from monolithic PostgreSQL to YugabyteDB\textquotesingle s distributed architecture, including

planning the migration, transforming schema, migrating data, and optimizing applications for a distributed environment. {[}11{]}{[}12{]}

There is also a useful tension inside YugabyteDB\textquotesingle s own public materials. A 2023 product blog says YugabyteDB 2.19\textquotesingle s bimodal query execution helps enable lift-and-shift migration of small and midsize applications. {[}13{]} That does not weaken the framework; it strengthens it. The same vendor can maintain a conservative docs-grade compatibility position while also marketing improved lift-and-shift capability for a bounded class of workloads. The institutional conclusion is still workload-specific: compatibility claims must be mapped to the behaviors the application actually depends on.

This is the key distinction:

PostgreSQL compatibility is not weak because vendors are overstating it. It is institutionally limited because the compatibility surface is deeper than the protocol.

Protocol, syntax, drivers, tools, and ORM support reduce adoption friction. They make evaluation cheaper. They let developers begin with familiar mental models. They may make many migrations practical.

They do not escrow production substitution unless the actual workload\textquotesingle s schema, extensions, planner assumptions, transaction behavior, operational tooling, backup and restore practices, performance envelope, and failure semantics have been validated.

The buyer\textquotesingle s rule should be:

Treat PostgreSQL compatibility as a reason to start an evaluation, not as a reason to end one.

\subsection{8.6 A Worked Assessment: S3-Compatible Storage for a Regulated Backup Workload}\label{a-worked-assessment-s3-compatible-storage-for-a-regulated-backup-workload}

The matrix assigns outcomes. A procurement decision needs one more step: scoring a specific claim against a specific workload. This section performs that step once, end to end, using only the public documentation already cited in Section 8.4. The example is deliberately the claim easiest to over-credit: S3 compatibility.

The scoring anchors below restate the five conditions as ordinal levels. They add no new theory; they make the conditions repeatable.

\begingroup\small
\begin{longtable}[]{@{}
  >{\raggedright\arraybackslash}p{(\columnwidth - 8\tabcolsep) * \real{0.19}}
  >{\raggedright\arraybackslash}p{(\columnwidth - 8\tabcolsep) * \real{0.27}}
  >{\raggedright\arraybackslash}p{(\columnwidth - 8\tabcolsep) * \real{0.27}}
  >{\raggedright\arraybackslash}p{(\columnwidth - 8\tabcolsep) * \real{0.27}}@{}}
\toprule\noalign{}
\begin{minipage}[b]{\linewidth}\raggedright
Condition
\end{minipage} & \begin{minipage}[b]{\linewidth}\raggedright
High
\end{minipage} & \begin{minipage}[b]{\linewidth}\raggedright
Medium
\end{minipage} & \begin{minipage}[b]{\linewidth}\raggedright
Low
\end{minipage} \\
\midrule\noalign{}
\endhead
\bottomrule\noalign{}
\endlastfoot
Boundary closure & The compatibility contract covers the production behaviors the workload relies on. & The core interaction model is covered; important perimeter behavior varies by implementation. & The claim covers protocol, syntax, or API shape only. \\
Executable conformance & Independent certification or a neutral conformance suite exists. & The buyer can run a repeatable workload-profile test. & Vendor documentation or sales assertion only. \\
Custody independence & The reference contract is governed outside the dominant vendor. & Governance is mixed or community-visible but not independent. & The dominant vendor controls the reference behavior. \\
State and operations reversibility & State, configuration, policy, and operating procedures move within a bounded program. & Data moves; policy, operations, and failure behavior need validation. & Migration requires workload-by-workload archaeology. \\
Extension quarantine & Non-portable features are inventoried and gated. & Departures can be profiled, but only manually. & No stable distinction between portable core and extensions. \\
\end{longtable}
\endgroup

The scenario: a regulated enterprise is evaluating whether an S3-compatible object store can serve as the backup and archive tier behind an existing object-storage dependency. The workload profile follows Section 8.4 --- required operations, headers, auth flows, bucket policies, lifecycle rules, event integrations, encryption requirements, object-lock semantics, replication expectations, failure behavior, performance assumptions, and retention controls --- with two additions the regulated context forces: auditability and restore performance.

Scoring the claim against that profile:

\begingroup\small
\begin{longtable}[]{@{}
  >{\raggedright\arraybackslash}p{(\columnwidth - 6\tabcolsep) * \real{0.22}}
  >{\raggedright\arraybackslash}p{(\columnwidth - 6\tabcolsep) * \real{0.60}}
  >{\raggedright\arraybackslash}p{(\columnwidth - 6\tabcolsep) * \real{0.18}}@{}}
\toprule\noalign{}
\begin{minipage}[b]{\linewidth}\raggedright
Condition
\end{minipage} & \begin{minipage}[b]{\linewidth}\raggedright
Assessment
\end{minipage} & \begin{minipage}[b]{\linewidth}\raggedright
Score
\end{minipage} \\
\midrule\noalign{}
\endhead
\bottomrule\noalign{}
\endlastfoot
Boundary closure & Strong object API center. Retention, object lock, lifecycle, IAM and bucket policy, replication, and eventing sit partly outside the stable core and vary by implementation. & Medium \\
Executable conformance & The profile can be tested repeatably against a target implementation, but the cited record establishes no neutral conformance regime comparable to OCI or Kubernetes. & Medium-low \\
Custody independence & The reference interface is controlled by AWS; third parties document support against it. & Low \\
State and operations reversibility & Object data is movable; retention, replication, policy, failure behavior, and restore procedures require validation before a bounded migration program can be claimed. & Medium \\
Extension quarantine & Departures can be recorded in a profile ledger, but nothing in the label forces the distinction; quarantine is available, not automatic. & Medium \\
\end{longtable}
\endgroup

Outcome: borrowed compatibility, with exit credit gated to the tested profile. The scores match the case matrix row for S3 compatibility in Section 8. What the worked pass adds is what the matrix cannot carry: a written record of which behaviors were tested, which were declared provider-specific extensions, and how much exit credit the claim earned.

The decision delta is visible in one clause. The untested version:

``Storage must be S3-compatible.''

The SET version:

``Storage must pass the defined S3 workload profile: required operations, headers, auth flows, bucket policies, lifecycle rules, event integrations, encryption requirements, object-lock semantics, replication expectations, failure behavior, performance assumptions, and retention controls. Unsupported or partially supported behavior is recorded as a provider-specific extension. Exit credit is limited to the tested profile.''

The first clause purchases a label. The second purchases a bounded substitution path. Nothing in the rewritten clause demands more compatibility from the vendor; it converts an unpriced adjective into a tested boundary, and it produces the inventory that extension quarantine requires.

\section{9. Similar Outcomes, Different Mechanisms}\label{similar-outcomes-different-mechanisms}

The framework matters because superficially similar cases are governed by different mechanisms.

OCI and S3 compatibility both reduce vendor exposure through widely implemented interfaces. OCI does it through escrow: independent custody, explicit specifications, and certification. S3 compatibility does it through ubiquity and implementation support against AWS\textquotesingle s reference API. Both are valuable. Only the first deserves general procurement-grade compatibility credit without workload profiling. {[}1{]}{[}2{]}{[}7{]}{[}8{]}{[}9{]}

Kubernetes and OpenTelemetry both reduce lock-in. Kubernetes escrows a core orchestration API while leaving much of the managed-service perimeter exposed. OpenTelemetry escrows instrumentation and telemetry export while leaving observability workflow exposed. Treating both as generic ``foundation portability'' loses the layer at which each actually works. {[}3{]}{[}4{]}{[}5{]}{[}6{]}

S3 compatibility and PostgreSQL compatibility both help teams avoid immediate application rewrites. But object storage has a simpler core state model than a production database. Database behavior is more semantic, workload-specific, and operationally entangled. Therefore the same phrase - compatible with the dominant interface - has different procurement meaning in the two categories.

Kubernetes conformance and PostgreSQL compatibility both involve large ecosystems and familiar APIs. Kubernetes conformance is executable against a defined certified surface. PostgreSQL compatibility in distributed databases is usually vendor documentation plus workload assessment. The difference is not community strength. It is whether the relevant production behavior sits inside an executable compatibility boundary. {[}3{]}{[}10{]}{[}11{]}{[}12{]}

These distinctions change decisions. A procurement team can accept OCI compatibility as a baseline requirement. It can require Kubernetes conformance but still demand a perimeter-migration assessment. It can require OpenTelemetry while still pricing observability-backend exit work. It can treat S3 compatibility as testable but not self-proving. It can treat PostgreSQL compatibility as migration acceleration rather than migration guarantee.

\section{10. Strategic Implications}\label{strategic-implications}

The first implication is that buyers should stop asking, ``Is it compatible?'' The correct question is:

Which future decision does this compatibility claim make cheaper?

OCI makes changing compliant registries, images, or runtimes cheaper inside the artifact/runtime boundary. Kubernetes conformance makes choosing among certified distributions and platforms safer at the API core. OpenTelemetry makes changing telemetry export paths cheaper at the instrumentation layer. S3 compatibility makes application rewrite less likely for tested object-storage operations. PostgreSQL compatibility makes database evaluation and partial migration easier, but production substitution remains workload-specific.

The second implication is that procurement should assign compatibility credit by layer, not by product.

A single platform may be portable at the artifact layer, standard at the orchestration layer, proprietary at the identity layer, non-portable at the data layer, and deeply locked in at the operational-workflow layer. A single ``open'' or ``compatible'' field in a risk spreadsheet destroys the useful information.

The third implication is that foundations should treat conformance as a product.

The institutional value of OCI and Kubernetes does not come only from specifications. It comes from evidence that can be attached to architecture review, procurement review, security admission, or risk assessment. A foundation that wants its project to matter in enterprise adoption should ask: what artifact can

the buyer put in the decision packet? If the answer is only a README, the project has not crossed the threshold.

The fourth implication is that vendors should decide where they want lock-in to live.

OpenTelemetry shifts lock-in downstream from service instrumentation into backend workflow. Kubernetes shifts some lock-in from orchestration API into managed-service perimeter. OCI shifts artifact lock-in out of individual container vendors and into surrounding platform, supply-chain, and operational choices. These are reallocations of dependency, not eliminations of dependency.

The fifth implication is that investors should not value compatibility uniformly.

A company building on OCI artifacts inherits a stronger compatibility substrate than a company claiming broad PostgreSQL substitutability. A storage company claiming S3 compatibility must prove the workload profile it covers. An observability vendor supporting OpenTelemetry may reduce ingestion friction while still competing through backend workflow differentiation. A distributed database claiming PostgreSQL compatibility may get faster evaluations but still needs workload-level proof.

The sixth implication is that security teams should use compatibility boundaries as control boundaries.

If the compatibility claim is executable, security can require evidence. If it is borrowed, security should require a workload profile and test results. If it is onboarding compatibility, security should treat it as unproven until production behavior is validated.

\section{11. What the Framework Predicts}\label{what-the-framework-predicts}

The Substitution Escrow Threshold makes auditable predictions.

First, OCI-compatible artifact requirements will continue to appear as baseline platform and supply-chain controls because the boundary is narrow, independently governed, and supported by certification evidence. Evidence that would weaken this prediction would be major enterprise platform standards treating OCI as insufficient even at the image/runtime boundary despite conformant tooling. {[}1{]}{[}2{]}

Second, certified Kubernetes will remain necessary but insufficient for migration risk reduction. The confirming signal is continued buyer diligence around storage, identity, networking, ingress, managed add-ons, upgrade procedures, observability, policy, and runbooks even when the vendor is CNCF-certified. The weakening signal would be routine enterprise migration between managed Kubernetes providers without perimeter remediation. {[}3{]}

Third, OpenTelemetry will reduce instrumentation lock-in more than observability-backend lock-in. The confirming signal is broad standardization of

OpenTelemetry emission while backend migrations still require work on dashboards, queries, alerts, retention, incident workflows, and historical data. The weakening signal would be repeated evidence of OpenTelemetry-standardized environments switching observability backends with minimal migration beyond endpoint changes. {[}4{]}{[}5{]}{[}6{]}

Fourth, S3-compatible procurement will become increasingly profile-based. The confirming signal is RFP and architecture-review language asking for supported operations, object-lock behavior, replication semantics, IAM or policy compatibility, lifecycle support, encryption modes, eventing behavior, and failure semantics rather than accepting ``S3-compatible'' as a binary claim. The weakening signal would be credible neutral conformance for S3-compatible implementations that large buyers accept without workload-specific testing. {[}7{]}{[}8{]}{[}9{]}

Fifth, PostgreSQL-compatible distributed databases will continue to win evaluations through ecosystem familiarity while still requiring workload-level certification for production substitution. The confirming signal is continued vendor investment in migration tools, unsupported-feature lists, extension support matrices, schema transformation guidance, and distributed-data-model documentation. The weakening signal would be repeated large-scale production migrations from PostgreSQL to distributed SQL systems without schema, query, extension, operational, or performance remediation. {[}10{]}{[}11{]}{[}12{]}

\section{12. Where the Framework Ends}\label{where-the-framework-ends}

The framework does not say buyers should always prefer escrowed compatibility. A proprietary system may be operationally superior. A managed service may reduce more risk than it creates. A single-vendor platform may have better support, security posture, cost profile, or integration depth.

The framework also does not apply where exit is irrelevant. If a component is disposable, compatibility escrow is unnecessary. If a platform\textquotesingle s value lies in proprietary data, workflow network effects, or managed operational excellence rather than substitutability, the framework is secondary.

Its purpose is narrower: to prevent institutions from mispricing compatibility claims.

A compatibility claim should receive procurement credit only for the future decision it actually makes cheaper.

The framework is also falsifiable at the framework level, beyond the per-claim weakening signals in Section 11. If claims scoring low on custody independence and executable conformance routinely deliver low-cost production substitution --- if S3-compatible stores substitute cleanly outside tested profiles, or PostgreSQL-compatible databases swap without workload remediation --- the five conditions are mispricing risk rather than measuring it. If claims scoring high on all five conditions still routinely fail to produce a bounded migration program, the threshold is set in the wrong place. And if new cases stop discriminating across outcome cells, the taxonomy is bookkeeping rather than a decision instrument.

\section{13. The Next Test: AI Compatibility Claims}\label{the-next-test-ai-compatibility-claims}

AI infrastructure is the framework\textquotesingle s most urgent next stress test: compatibility claims there are multiplying faster than conformance evidence. The Substitution Escrow Threshold applies without modification.

The claims are already live: vLLM\textquotesingle s OpenAI compatible server, Ollama\textquotesingle s partial OpenAI API compatibility, LiteLLM\textquotesingle s OpenAI-compatible gateway, OpenRouter\textquotesingle s provider-normalized OpenAI Chat API surface, Anthropic\textquotesingle s Model Context Protocol, vector-database API surfaces around Pinecone, Weaviate, Qdrant, and Milvus, and model-serving or evaluation harness compatibility claims that present themselves as future substitution paths. {[}14{]}{[}15{]}{[}16{]}{[}17{]}{[}18{]}{[}21{]}{[}22{]}{[}23{]}{[}24{]}

The institutional anchor is already visible. CNCF announced the Certified Kubernetes AI Conformance Program on November 11, 2025 at KubeCon + CloudNativeCon North America in Atlanta. The program is a community-led effort to define and validate standards for AI workloads on Kubernetes, had certified initial participants with a v1.0 release, and began work on a v2.0 roadmap. The public project repository describes the program as defining the capabilities a Kubernetes platform needs to run AI and machine-learning workloads reliably, with certification based on reviewable evidence today and automated conformance tests planned for 2026. {[}19{]}{[}20{]}

Those cases cover the same decision surface as the five infrastructure examples above: which compatibility claims actually escrow future substitution, and which merely reduce first-endpoint integration cost? OpenAI-shaped HTTP endpoints may reduce application rewrites. MCP may escrow part of the agent tool integration boundary. Kubernetes AI conformance may escrow part of the workload-platform boundary. Vector-database APIs may reduce client integration friction. None of those claims should be treated as production substitutability until the boundary, conformance evidence, custody, state reversibility, and extension drift have been tested.

Against those candidates, the framework predicts that many AI compatibility claims will initially sit in borrowed or onboarding-compatible cells. Endpoint mimicry is easy. Behavioral substitutability is hard. Model behavior, latency distributions, context handling, tool-calling semantics, structured-output reliability, safety filters, eval harnesses, cost envelopes, data-retention commitments, and incident procedures are not captured by an HTTP shape alone. Each of those assignments is scoreable today with the five conditions and the anchors of Section 8.6; no new machinery is required.

That does not make the claims useless. It means they should be priced correctly. Compatibility is not a virtue. It is a claim about a future institutional decision.

The only claims that deserve procurement credit are the ones that put enough of that future decision into escrow.

\section{References}\label{references}

{[}1{]} Open Container Initiative. ``About the Open Container Initiative.'' \url{https://opencontainers.org/about/overview/}

{[}2{]} Open Container Initiative. ``OCI Certified.'' \url{https://opencontainers.org/community/certified/}

{[}3{]} Cloud Native Computing Foundation. ``Certified Kubernetes Software Conformance.'' \url{https://www.cncf.io/training/certification/software-conformance/}

{[}4{]} OpenTelemetry. ``Documentation.'' \url{https://opentelemetry.io/docs/} 

{[}5{]} OpenTelemetry. ``Collector.'' \url{https://opentelemetry.io/docs/collector/}

{[}6{]} Cloud Native Computing Foundation. ``OpenTelemetry.'' \url{https://www.cncf.io/projects/opentelemetry/}

{[}7{]} AWS Documentation. ``S3 API Reference - Amazon Simple Storage Service.'' \url{https://docs.aws.amazon.com/AmazonS3/latest/API/Welcome.html}

{[}8{]} Ceph Documentation. ``Ceph Object Gateway S3 API.'' \url{https://docs.ceph.com/en/latest/radosgw/s3/}

{[}9{]} MinIO Documentation. ``S3 API Compatibility.'' \url{https://docs.min.io/aistor/developers/s3-api-compatibility/}

{[}10{]} Cockroach Labs Documentation. ``PostgreSQL Compatibility.'' \url{https://www.cockroachlabs.com/docs/stable/postgresql-compatibility}

{[}11{]} YugabyteDB Documentation. ``FAQs about YugabyteDB API compatibility.'' \url{https://docs.yugabyte.com/stable/faq/compatibility/}

{[}12{]} YugabyteDB Documentation. ``Migrate from PostgreSQL.'' \url{https://docs.yugabyte.com/stable/manage/data-migration/migrate-from-postgres/}

{[}13{]} YugabyteDB. ``Dream Big, Go Bigger: Turbocharging PostgreSQL.'' \url{https://www.yugabyte.com/blog/postgresql-turbocharging/}

{[}14{]} vLLM Documentation. ``Online Serving / OpenAI-Compatible Server.'' \url{https://docs.vllm.ai/en/latest/serving/online_serving/}

{[}15{]} Ollama Documentation. ``OpenAI compatibility.'' \url{https://docs.ollama.com/api/openai-compatibility}

{[}16{]} LiteLLM Documentation. ``Getting Started.'' \url{https://docs.litellm.ai/docs/}

{[}17{]} OpenRouter Documentation. ``API Reference.'' \url{https://openrouter.ai/docs/api/reference/overview}

{[}18{]} Model Context Protocol Documentation. ``What is the Model Context Protocol?'' \url{https://modelcontextprotocol.io/docs/getting-started/intro}

{[}19{]} Cloud Native Computing Foundation. ``CNCF Launches Certified Kubernetes AI Conformance Program to Standardize AI Workloads on Kubernetes.'' \url{https://www.cncf.io/announcements/2025/11/11/cncf-launches-certified-kubernetes-ai-conformance-program-to-standardize-ai-workloads-on-kubernetes/}

{[}20{]} CNCF GitHub. ``Kubernetes AI Conformance.'' \url{https://github.com/cncf/k8s-ai-conformance}

{[}21{]} Pinecone Documentation. ``API reference.'' \url{https://docs.pinecone.io/reference/api/introduction}

{[}22{]} Weaviate Documentation. ``APIs.'' \url{https://docs.weaviate.io/weaviate/api}

{[}23{]} Qdrant Documentation. ``API \& SDKs.'' \url{https://qdrant.tech/documentation/interfaces/}

{[}24{]} Milvus Documentation. ``About RESTful API.'' \url{https://milvus.io/api-reference/restful/v2.4.x/About.md}

{[}25{]} Farrell, J. and Saloner, G. ``Standardization, Compatibility, and Innovation.'' RAND Journal of Economics 16(1), 1985, pp. 70--83.

{[}26{]} Katz, M. L. and Shapiro, C. ``Network Externalities, Competition, and Compatibility.'' American Economic Review 75(3), 1985, pp. 424--440.

{[}27{]} David, P. A. ``Clio and the Economics of QWERTY.'' American Economic Review 75(2), 1985, pp. 332--337.

{[}28{]} Arthur, W. B. ``Competing Technologies, Increasing Returns, and Lock-in by Historical Events.'' Economic Journal 99(394), 1989, pp. 116--131.

{[}29{]} Liebowitz, S. J. and Margolis, S. E. ``Path Dependence, Lock-in, and History.'' Journal of Law, Economics, and Organization 11(1), 1995, pp. 205--226.

{[}30{]} David, P. A. and Greenstein, S. ``The Economics of Compatibility Standards: An Introduction to Recent Research.'' Economics of Innovation and New Technology 1(1--2), 1990, pp. 3--41.

{[}31{]} Besen, S. M. and Farrell, J. ``Choosing How to Compete: Strategies and Tactics in Standardization.'' Journal of Economic Perspectives 8(2), 1994, pp. 117--131.

{[}32{]} Shapiro, C. and Varian, H. R. Information Rules: A Strategic Guide to the Network Economy. Harvard Business School Press, 1999.

{[}33{]} Farrell, J. and Klemperer, P. ``Coordination and Lock-In: Competition with Switching Costs and Network Effects.'' In Handbook of Industrial Organization, Vol. 3, ed. M. Armstrong and R. Porter, Elsevier, 2007, pp. 1967--2072.

{[}34{]} Raemaekers, S., van Deursen, A., and Visser, J. ``Semantic versioning and impact of breaking changes in the Maven repository.'' Journal of Systems and Software 129, 2017, pp. 140--158.

{[}35{]} Decan, A. and Mens, T. ``What Do Package Dependencies Tell Us About Semantic Versioning?'' IEEE Transactions on Software Engineering 47(6), 2021, pp. 1226--1240.

{[}36{]} Lacity, M. C., Khan, S. A., Yan, A., and Willcocks, L. P. ``A Review of the IT Outsourcing Empirical Literature and Future Research Directions.'' Journal of Information Technology 25(4), 2010, pp. 395--433.

{[}37{]} Opara-Martins, J., Sahandi, R., and Tian, F. ``Critical analysis of vendor lock-in and its impact on cloud computing migration: a business perspective.'' Journal of Cloud Computing: Advances, Systems and Applications 5, 2016, Article 4.

{[}38{]} Benaroch, M. ``Managing Information Technology Investment Risk: A Real Options Perspective.'' Journal of Management Information Systems 19(2), 2002, pp. 43--84.

{[}39{]} Fichman, R. G. ``Real Options and IT Platform Adoption: Implications for Theory and Practice.'' Information Systems Research 15(2), 2004, pp. 132--154.

{[}40{]} Intellectual Property Bankruptcy Protection Act of 1988, 11 U.S.C. § 365(n).

{[}41{]} ISO/IEC 19941:2017. Information technology --- Cloud computing --- Interoperability and portability. International Organization for Standardization and International Electrotechnical Commission, 2017.

{[}42{]} Regulation (EU) 2016/679 of the European Parliament and of the Council of 27 April 2016 (General Data Protection Regulation), Article 20 (Right to data portability).

{[}43{]} Regulation (EU) 2023/2854 of the European Parliament and of the Council of 13 December 2023 on harmonised rules on fair access to and use of data (Data Act), Chapter VI (Articles 23--31).

\end{document}